\documentclass[conference]{IEEEtran}
\IEEEoverridecommandlockouts

\usepackage[utf8]{inputenc}
\usepackage[T1]{fontenc}
\usepackage{comment}
\usepackage{cite}
\usepackage{amsmath,amssymb,amsfonts}
\usepackage{algorithmic}
\usepackage[linesnumbered,ruled]{algorithm2e}
\usepackage{graphicx}
\usepackage{textcomp}
\usepackage{xcolor}
\usepackage{booktabs}
\usepackage{listings}
\usepackage{mdframed}
\usepackage{cleveref}
\usepackage{breqn}
\usepackage{stfloats}
\usepackage{afterpage}
\usepackage{float}
\usepackage{here}
\usepackage{subcaption}
\usepackage{url} 
\usepackage{svg}
\usepackage{tikz}

\usepackage[most]{tcolorbox} 
\def\BibTeX{{\rm B\kern-.05em{\sc i\kern-.025em b}\kern-.08em
    T\kern-.1667em\lower.7ex\hbox{E}\kern-.125emX}}

\crefname{figure}{Fig.}{Figs.}
\crefname{table}{TABLE}{Tables}
\crefname{section}{Section}{Sections}
\crefname{equation}{eq.}{eqs.}

\begin{document}

\newcommand{\AGNOCAST}{\textit{Agnocast}}
\newcommand{\SMARTPOINTER}{\textit{ipc\_shared\_ptr}}

\newcommand{\ieeecopyrightfooter}{%
  \begin{tikzpicture}[remember picture,overlay]
    \node[anchor=south,yshift=0.5cm,
          text width=\dimexpr\textwidth\relax,
          align=left,font=\footnotesize]
      at (current page.south)
      {\textcopyright\ 2026 IEEE. Personal use of this material is permitted. Permission from IEEE must be obtained for all other uses, in any current or future media, including reprinting/republishing this material for advertising or promotional purposes, creating new collective works, for resale or redistribution to servers or lists, or reuse of any copyrighted component of this work in other works. Accepted for publication in the 2026 IEEE 29th International Symposium on Real-Time Distributed Computing (ISORC).};
  \end{tikzpicture}%
}

\title{
ipc\_shared\_ptr: A Publish/Subscribe-Aware\\Smart Pointer for Cross-Process Object Lifetime Management
}

\author{
  \IEEEauthorblockN{%
    Takahiro Ishikawa-Aso$^{\dagger\ddagger}$,
    Atsushi Yano$^{*\ddagger}$,
    Koichi Imai$^{\ddagger}$,
    Takuya Azumi$^{*}$,
    Shinpei Kato$^{\dagger}$}
  \IEEEauthorblockA{%
    $^{\dagger}$The University of Tokyo, Japan \quad
    $^{*}$Saitama University, Japan \quad
    $^{\ddagger}$TIER IV Incorporated, Japan}
}

\lstdefinestyle{mystyle}{
    basicstyle=\ttfamily\footnotesize,
    breakatwhitespace=false,
    breaklines=true,
    captionpos=b,
    keepspaces=true,
    numbers=left,
    numbersep=10pt, 
    showspaces=false,
    showstringspaces=false,
    showtabs=false,
    tabsize=2,
    framexleftmargin=15pt, 
    xleftmargin=15pt,
    morekeywords={agnocastlib, LD\_PRELOAD, MEMPOOL\_SIZE},
    keywordstyle=\color{red}\bfseries, 
}

\lstset{style=mystyle}

\newtcolorbox{codebox}[1][]{
    enhanced,
    unbreakable,
    colback=white,
    colframe=black,
    boxrule=0.5pt,
    sharp corners,
    boxsep=2pt,
    left=5pt,
    right=5pt,
    top=2pt,
    bottom=2pt,
    listing only,
    listing options={style=mystyle},
    overlay={\node[anchor=south east, outer sep=2pt, font=\footnotesize] at ([xshift=-3pt, yshift=3pt]frame.south east) {#1};}
}

\maketitle
\ieeecopyrightfooter

\begin{abstract}
True zero-copy Inter-Process Communication (IPC) in
publish/subscribe (pub/sub) middleware such as Robot Operating System 2 (ROS~2) requires
subscribers to reference message objects in publisher-owned shared
memory.
Objects must not be reclaimed while referenced, yet must eventually be reclaimed, with correct handling of crash recovery and Transient Local QoS retention requirements.
We propose \SMARTPOINTER{}, a pub/sub-aware smart pointer for
cross-process message lifetime management.
\SMARTPOINTER{} exploits pub/sub structural properties to specialize
Birrell's reference listing, limiting global metadata updates to
per-subscriber 0$\leftrightarrow$1 transitions and achieving an
order-of-magnitude reduction in global communication over
general-purpose distributed reference counting.
We analyze the key metadata management tradeoff: scalability versus
implementation simplicity. Owner-driven reclaim offers greater scalability, but concurrent membership changes and reclamation decisions produce races that widen the correctness-verification state space.
Single-writer achieves structural atomicity, eliminating this
complexity at the cost of a centralized bottleneck.
iceoryx2 (owner-driven reclaim) and Agnocast --- a true zero-copy
ROS~2 IPC middleware sharing the publisher's heap with subscribers
and adopting \SMARTPOINTER{} with single-writer --- embody each
architecture.
Comparative evaluation at the scale of Autoware --- the largest open-source ROS~2 application --- confirms that single-writer achieves sufficient scalability: at 200 topics, two subscribers per topic and 100~Hz, Agnocast's E2E p99.9 is 2.9$\times$ lower than iceoryx2's, justifying implementation simplicity over owner-driven reclaim.
\end{abstract}


\vspace{-5mm}

\section{Introduction} \label{section-introduction}
Many autonomous cyber-physical systems such as autonomous driving and robotics systems are built as component-oriented real-time systems \cite{cie2025wip}, where independent nodes exchange messages through publish/subscribe (pub/sub) \cite{eugster2003many} middleware such as Robot Operating System 2 (ROS~2) \cite{ros2doc}.
When nodes run as separate processes for fault isolation, Inter Process Communication (IPC) cost directly impacts system performance.
True zero-copy communication eliminates all copying including serialization and deserialization, reducing transfer latency, freeing CPU cycles, avoiding temporary memory allocations, and improving timing predictability.

In true zero-copy IPC, subscriber processes directly reference message objects constructed on publisher-owned shared memory.
iceoryx2 \cite{iceoryx2} achieves true zero-copy IPC with pool-allocated chunks, restricting payloads to self-contained, trivially destructible types while allowing per-message metadata to be embedded within each chunk.
Agnocast \cite{agnocast2025} instead maps the publisher's heap to shared memory, supporting arbitrary message types including those with heap allocations such as \lstinline|std::vector|; however, because subscribers see the publisher's heap as-is, there is no reserved space for per-message metadata.
Agnocast therefore requires an external mechanism for cross-process object lifetime management.

Cross-process lifetime management is non-trivial.
Process-local reference counts (as in C++'s \lstinline|std::shared_ptr|, which deallocates an object when its in-process reference count reaches zero) are insufficient: whether a message can be reclaimed depends on references held across all processes.
Dynamic membership --- joins, leaves, and crashes --- and Transient Local Quality of Service (QoS), which retains messages for late joiners, add further complexity.
A naive global per-message reference count is costly because every reference copy, whether intra- or inter-process, requires a synchronized update.
The distributed systems literature has long studied two-level approaches separating process-local from global reference management to reduce coordination costs \cite{plainfosse1995survey}; we adapt such schemes to publish/subscribe systems.
In particular, Birrell's reference listing \cite{birrell1993distributed} is the closest antecedent. 
In that scheme, the object owner maintains process IDs holding references, limiting global communication to process-granularity transitions.
However, such general-purpose methods assume references can propagate between arbitrary processes.
In pub/sub, the topic graph determines which processes receive each message, enabling a simpler, more efficient design.

This paper proposes \SMARTPOINTER{} for cross-process object lifetime management in pub/sub systems, exploiting three structural properties: (1) the reference holder set is derivable a priori from topic
membership, (2) references flow unidirectionally from publishers to subscribers, and (3) messages are independent deallocation units.
The result specializes Birrell's reference listing to the pub/sub domain: global metadata updates are limited to 0$\leftrightarrow$1 transitions of each subscriber's local reference count; in-process copies trigger no global updates.
In robotics workloads, where references to a message are copied multiple times within a single process while total process counts are limited, this achieves an order-of-magnitude reduction in global communication over general-purpose methods.

Cross-process lifetime management requires per-message metadata accessible to all participating processes, whose placement creates a key tradeoff between scalability and implementation simplicity.
In the \emph{single-writer} approach, a centralized server mediates all metadata access, structurally guaranteeing atomicity.
In the \emph{owner-driven reclaim} approach adopted by iceoryx2 \cite{iceoryx2}, each publisher manages its own metadata without centralized coordination, offering better theoretical scalability at the cost of races between concurrent membership changes and reclamation decisions (\cref{section-tradeoffs}).
Agnocast adopts single-writer; comparative evaluation with iceoryx2 shows it scales to Autoware-class workloads \cite{kato2015open, kato2018autoware}, justifying implementation simplicity over theoretical scalability.

\textbf{Contributions.}
(1)~We propose \SMARTPOINTER{}, a cross-process object lifetime management mechanism for pub/sub systems. By specializing Birrell's reference listing to pub/sub, \SMARTPOINTER{} derives the receiver set from topic membership instead of registering references on each transfer, and bounds the global reference set to current subscribers. We demonstrate feasibility on Agnocast.
(2)~We comparatively evaluate the single-writer (Agnocast) and owner-driven reclaim (iceoryx2) architectures, and show that the single-writer design achieves comparable scalability at Autoware-class scale.

\section{Background} \label{section-background}

\subsection{Publish/Subscribe Systems}
\vspace{-1mm}

\textbf{Topic-based messaging.}
Publish/subscribe is a messaging paradigm in which publishers and subscribers communicate through named topics \cite{eugster2003many}. Publishers send messages without knowledge of which subscribers exist; all current subscribers of that topic receive each published message.
A key characteristic is \textit{dynamic membership}: throughout this paper, the \textit{membership} of a topic refers to the set of publishers and subscribers currently participating in that topic, which changes as endpoints join, leave, or crash at any time.

\textbf{Quality of Service (QoS).}
Two QoS parameters \cite{ros2qosdoc} are directly relevant to object lifetime management.
\emph{Durability} determines whether messages are retained for late-joining subscribers: under Transient Local, the publisher retains published messages and delivers history to late joiners, whereas under Volatile, messages are delivered only to subscribers present at publish time.
\emph{Depth}, under the Keep Last policy, specifies how many past messages the publisher retains for Transient Local delivery.
Even after all current subscribers release their references, the publisher must retain up to Depth messages for future late joiners --- directly affecting when message objects can be reclaimed (\cref{section-ipc-shared-ptr}).

\begin{figure}[tb]
  \centering
  \includegraphics[width=\linewidth]{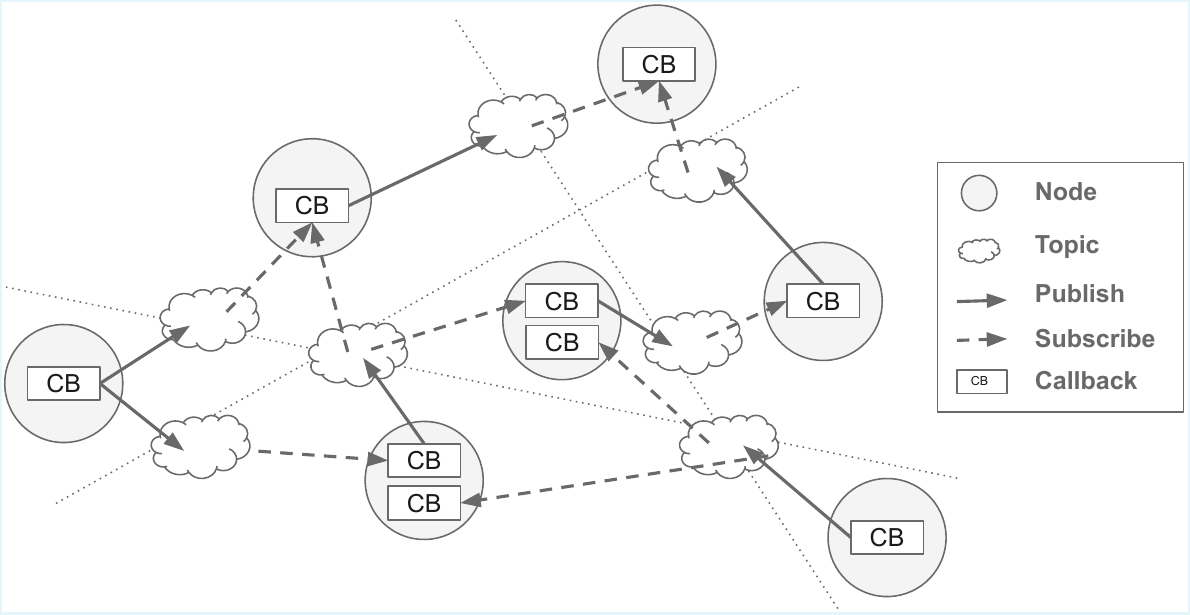}
  \vspace{-5mm}
  \caption{Publish/subscribe system (Robot Operating System 2).}
  \label{fig-pubsub-system}
  \vspace{-6mm}
\end{figure}

\subsection{Robot Operating System 2}
\vspace{-1mm}

\textbf{Overview.}
ROS~2 \cite{ros2doc} is the de facto development platform for
robotics and autonomous driving, adopting a component-oriented
architecture where independent nodes communicate through
topic-based pub/sub.
Each topic is identified by a name and associated with a
message type.
Each node contains one or more callbacks, which are triggered upon
message reception; a callback may in turn publish messages to other
topics, forming system-wide dataflows
(\cref{fig-pubsub-system}).

\textbf{Message types and rosidl.}
Message types are defined in interface definition files (\texttt{.msg}) and compiled into language-specific data structures by the rosidl toolchain \cite{ros2rosidldoc}.
Message types containing dynamic containers such as \lstinline|std::vector| may undergo reallocation during message construction; this paper refers to these as \emph{unsized message types}.
Nearly all standard ROS~2 message packages (\texttt{std\_msgs}, \texttt{sensor\_msgs}, \texttt{geometry\_msgs}, etc.) contain unsized types \cite{agnocast2025}.

\textbf{Allocator constraint and Agnocast.}
True zero-copy IPC requires heap allocations to be redirected to shared memory. In ROS~2 C++, rosidl-generated structs hardcode \lstinline|std::allocator<void>| for all container members, with no mechanism to parameterize the allocator type \cite{ros2rosidldoc}. Three approaches have been considered, each breaking type compatibility: (1)~a custom allocator changes the type signature; (2)~a polymorphic allocator resolves the template parameter but breaks container compatibility; (3)~making unsized fields implicitly bounded preserves the API but breaks type consistency \cite{rclcpp2201}. See \cite{agnocast2025} for details.
Agnocast sidesteps this by mapping the publisher's heap itself to shared memory at an identical virtual address offset across publisher and subscriber processes, so that standard rosidl-generated types can be used unmodified~\cite{agnocast2025}.

\textbf{Autoware.}
Autoware \cite{kato2015open, kato2018autoware} is the largest open-source autonomous driving stack built on ROS~2. Autoware v1.7.1 \cite{autoware171} comprises 223 nodes and 227 topics\footnote{Counting only topics with at least one publisher and one subscriber.}, and serves as the workload baseline in \cref{section-evaluation}. The median number of subscribers per topic is 1 (mean 3.4); the five most subscribed topics are vehicle pose estimate~(35), coordinate transforms~(32), static coordinate transforms~(31), HD map~(25), and planned route~(16). LiDAR point clouds drive the majority of the pipeline at 10~Hz; exceptions include vehicle control commands at 33~Hz and localization/velocity outputs at 50--100~Hz. Camera inputs operate at 10~Hz, radar at 20~Hz. Point cloud messages span several megabytes per frame, simultaneously consumed by multiple nodes for detection, localization, and planning.

\section{Problem Statement} \label{section-problem-statement}
In true zero-copy IPC, subscribers directly reference
message objects on publisher-owned shared memory, raising the
question of when and how to reclaim them.
This section formalizes the system model (\cref{fig-system-model}) and requirements.

\subsection{System Model} \label{section-system-model}
The target is single-host multi-process pub/sub; network-distributed environments are out of scope.
Subscribers directly reference message objects in publisher-owned shared memory.
Processes have separate address spaces and may start, terminate, or crash at any time.
Each message is an independent deallocation unit with no inter-object references; all ROS~2 message types satisfy this property.
Subscriber processes access these objects in read-only mode, holding references in arbitrary number and releasing them at any time.

\begin{figure}[tb]
  \centering
  \includegraphics[width=\linewidth]{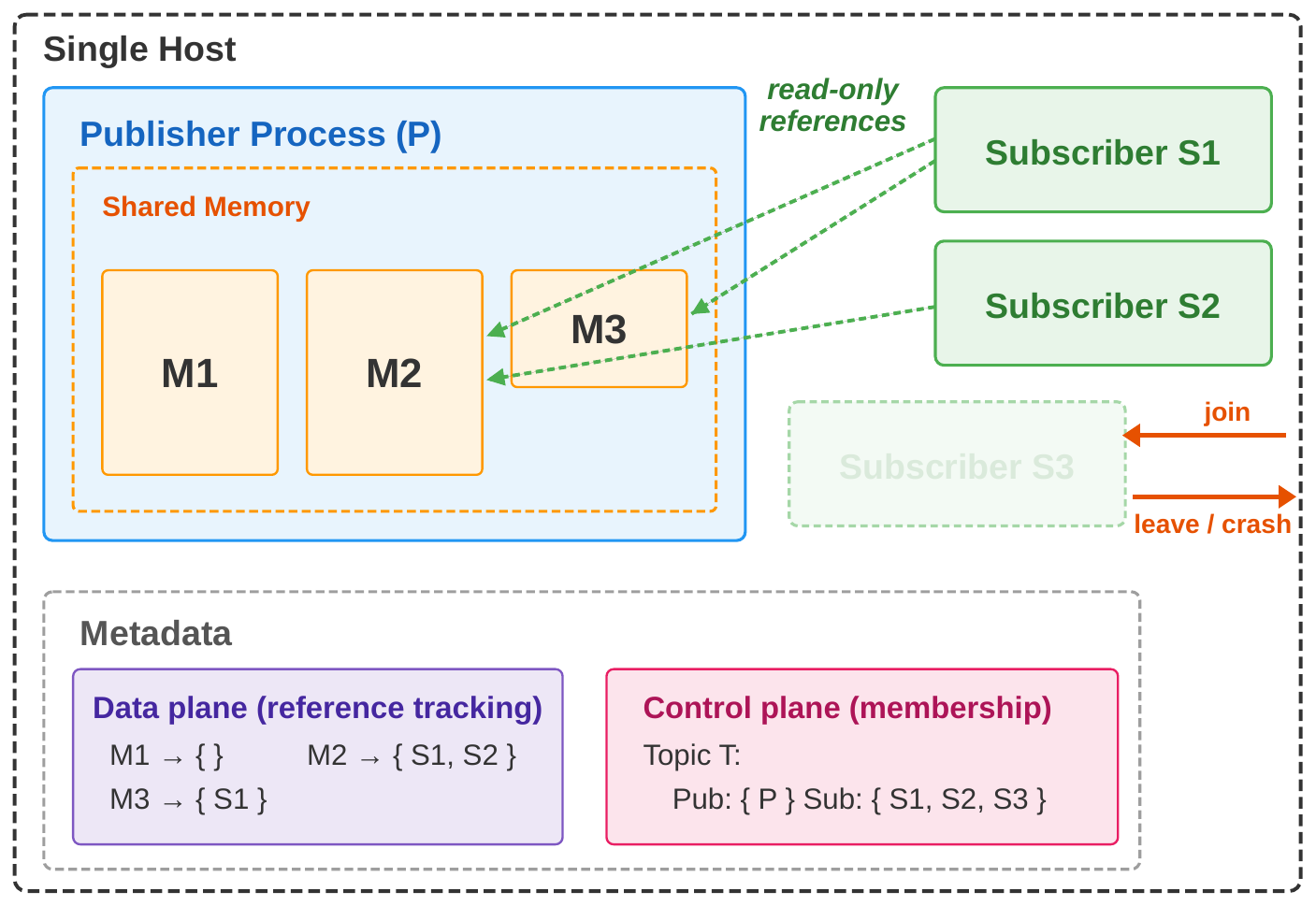}
  \vspace{-5mm}
  \caption{System model behind \SMARTPOINTER{}.}
  \label{fig-system-model}
  \vspace{-5mm}
\end{figure}

This paper classifies metadata into two categories: the \emph{data plane}, which tracks per-message reference holders and reclamation eligibility; and the \emph{control plane}, which manages topic membership, membership change timing, and late-joiner catch-up positions.
In \cref{fig-system-model}, the data plane records that M1 has no remaining references and is reclaimable, while M2 is still referenced by S1 and S2; the control plane records the membership of topic~T and the position from which a late joiner such as S3 should begin receiving messages.

\subsection{Requirements} \label{section-requirements}

A cross-process lifetime management mechanism must satisfy the
following requirements.

\textbf{R1 (No premature reclamation).}
Objects must not be reclaimed while any subscriber holds a
reference.
With Transient Local QoS, reclamation must additionally be
suppressed until Depth retention requirements are met, even when
all subscriber references have been released.

\textbf{R2 (No permanent leak).}
Objects whose references have all been released and whose
retention requirements are met must eventually be reclaimed.

\textbf{R3 (Transparency).} The complexity of cross-process lifetime management must be hidden from application code, which must be able to use \SMARTPOINTER{} with \lstinline|std::shared_ptr|-equivalent copy, move, and destruction semantics.

\textbf{R4 (Dynamic membership).}
R1--R3 must hold across membership changes, including unexpected
crashes.
Orphaned references from crashed processes must be cleaned up
without violating R1.
History delivery to late joiners must execute safely in parallel
with reclamation and crash recovery.

\textbf{R5 (Scalability).}
Metadata management overhead must not substantively degrade performance at Autoware-class scale.

R1--R5 apply regardless of the metadata management architecture.
The key difference between architectures is the implementation
complexity required to achieve them.
\cref{section-metadata-architecture} shows that
data plane/control plane separation increases the protocol
complexity needed for R1, R2 and R4;
\cref{section-evaluation} shows that the single-writer
approach satisfies R5.
Together, these support adopting the single-writer approach.

\section{\SMARTPOINTER{}} \label{section-ipc-shared-ptr}
\vspace{-1mm}

\SMARTPOINTER{} manages cross-process message lifetimes in pub/sub systems via two-level reference counting: process-local counts and membership-aware global tracking.
The two-level structure eliminates the kernel call per reference copy incurred by naive single-level designs such as pre-v2.2.0 Agnocast (\cref{section-birrell-relationship}).
The mechanism is independent of the shared memory design (heap-sharing or chunk-based).
\cref{section-implementation} demonstrates \SMARTPOINTER{} on Agnocast.

\vspace{-1mm}
\subsection{Design} \label{section-definition}
\begin{figure}[tb]
  \centering
  \includegraphics[width=\linewidth]{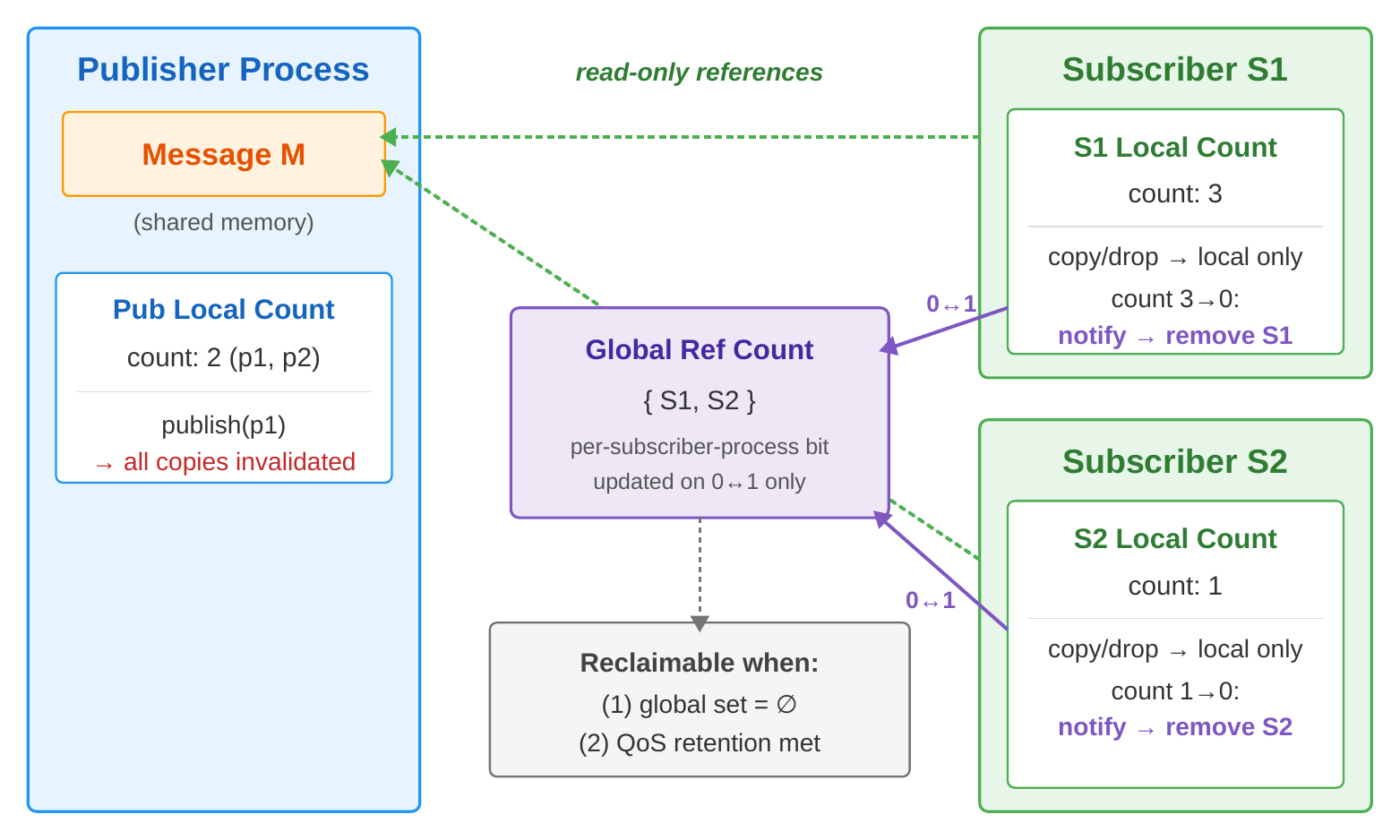}
  \vspace{-5mm}
\caption{Two-level reference count structure of \SMARTPOINTER{}. \lstinline|p1|, \lstinline|p2| denote distinct \SMARTPOINTER{} copies in the publisher process to the same message \lstinline|M|; \lstinline|publish(p1)| publishes \lstinline|M| and invalidates all such copies. Of the three counts, only the global reference count belongs to the data plane in \cref{fig-system-model}; pub/sub local counts reside within each process.}
  \label{fig-definition}
  \vspace{-5mm}
\end{figure}

\SMARTPOINTER{} combines a two-level reference count structure
with pub/sub-specific ownership semantics (\cref{fig-definition}).

\textbf{Publisher-side semantics.}
A publisher constructs a message via \SMARTPOINTER{} with \lstinline|std::shared_ptr|-equivalent copy, move, and destruction semantics; copies within the publisher process (e.g., \lstinline|p1|, \lstinline|p2| in \cref{fig-definition}) are tracked by a publisher's local count.
\lstinline|publish()| transfers ownership to the middleware; all copies the publisher still holds are invalidated, with subsequent accesses detected as runtime errors. This prevents writes to data being simultaneously read by subscribers, catching bugs during CI/CD and simulation testing.
An \SMARTPOINTER{} whose local count reaches zero without \lstinline|publish()| deallocates exactly like \lstinline|std::shared_ptr|.\footnotemark{}

\afterpage{\footnotetext{A \lstinline|std::unique_ptr|-equivalent design is possible, precluding copies before publish and removing the need for invalidation. The Agnocast implementation adopts \lstinline|std::shared_ptr| semantics because the existing ROS~2 API permits referencing a message from multiple locations before publish (e.g., publishing the same message to multiple topics), and ROS~2 developers are accustomed to this interface.}}

\textbf{Subscriber-side semantics.}
A subscriber accesses each received message through an \SMARTPOINTER{} that behaves as a \lstinline|std::shared_ptr|: copy, move, and destruction semantics are identical from the subscriber's perspective.
In-process copies increment only the process-local count; no
cross-process communication occurs.
When the local count drops to zero, the subscriber notifies the
metadata manager (via syscall in the kernel module implementation)
to remove itself from the global reference set, rather than
freeing the object.

\textbf{Global reference count.}
A per-message set of subscriber process IDs, each indicating that
the process holds at least one local reference.
An entry is added when a subscriber first acquires a reference
(at publish time or late-joiner delivery) and removed upon the
subscriber's zero-count notification.
Only 0$\leftrightarrow$1 transitions of the local count trigger
global updates; intermediate increments and decrements are
invisible globally.
The management method for the global reference set --- centralized
versus distributed --- is the subject of
\cref{section-metadata-architecture}.

\textbf{Reclaimable timing.}
A message becomes eligible for reclamation when two conditions are
jointly satisfied:
(1)~the global reference set is empty, and
(2)~QoS retention requirements are met --- with Transient Local
QoS, at least Depth newer messages from the same publisher exist.
Actual deallocation occurs at an arbitrary time after both
conditions hold.
When and by whom the reclamation check is performed depends on the
metadata architecture (\cref{section-metadata-architecture}).

\subsection{Relationship to Birrell's Reference Listing}
\label{section-birrell-relationship}

\SMARTPOINTER{} specializes Birrell et al.'s reference listing
\cite{birrell1993distributed}, the closest antecedent in the
distributed reference counting literature
\cite{plainfosse1995survey}.
In Birrell's method, the object owner maintains a set of
referencing process IDs (\emph{dirtySet}); a client issues a
\emph{dirty Remote Procedure Call (RPC)} on first surrogate creation and a \emph{clean
RPC} when the surrogate becomes locally unreachable.
This ``process-local tracking + owner-managed ID set'' pattern
maps directly to the two-level structure in
\cref{section-definition}.

\SMARTPOINTER{} exploits three structural properties of pub/sub
to simplify this pattern:

\textbf{A priori reference set.}
Birrell requires a dirty RPC per transfer to each new process,
since references can propagate between arbitrary process pairs.
In pub/sub, topic membership fully determines the receiver set;
references do not propagate between subscribers.
Dirty RPCs are thus absorbed into publish and membership
operations.

\textbf{Bounded global update frequency.}
Because the reference set is known a priori, a
per-subscriber-process bit suffices --- set on first local
reference, cleared on last.
Global updates occur only on 0$\leftrightarrow$1 transitions,
yielding $O(\text{subscribers})$ frequency rather than
$O(\text{reference copies})$.

\textbf{No circular references.}
Messages are independent deallocation units with no inter-object
references, structurally eliminating the circular reference
problem left unresolved in Birrell's work.

\begin{table}[tb]
\centering
\caption{Per-message global metadata update frequency.}
\label{tab-communication-cost}
\setlength{\tabcolsep}{2pt}
\resizebox{\columnwidth}{!}{
\begin{tabular}{@{}lll@{}}
\toprule
Method & Updates per message & Mechanism \\
\midrule
Distributed ref.\ counting
  & $O(\text{ref.\ copies})$
  & Per-copy global update \\[3pt]
Weighted ref.\ counting \cite{bevan1987distributed, watson1987efficient}
  & Amort.\ $O(1)$
  & Weight redistrib. \\[3pt]
Ref.\ listing \cite{birrell1993distributed}
  & $O(\text{proc.\ transitions})$
  & Dirty/clean RPCs \\[3pt]
\SMARTPOINTER{}
  & $O(\text{subscribers})$
  & Publish/membership ops \\
\bottomrule
\end{tabular}
} 
\vspace{-4mm}
\end{table}

\cref{tab-communication-cost} compares per-message global update
frequency across methods.
Weighted reference counting \cite{bevan1987distributed, watson1987efficient} achieves
amortized $O(1)$ by splitting a finite weight among references,
but requires a fallback protocol when the weight is exhausted ---
unnecessary in pub/sub where the receiver set is bounded by membership.
Reference listing shares the same asymptotic order as
\SMARTPOINTER{}, but requires explicit dirty/clean RPCs per
transfer; \SMARTPOINTER{} absorbs these into existing
publish and membership operations.
The practical gap is large in ROS~2: messages are typically copied
across multiple callbacks within each subscriber, while subscriber
counts remain small (median 1, mean 3.4 in Autoware).

\section{Metadata Management Architecture}
\label{section-metadata-architecture}

The data plane and control plane (\cref{section-system-model}) can be managed anywhere from fully centralized to fully distributed. This section examines two representative extremes --- single-writer and owner-driven reclaim (iceoryx2 \cite{iceoryx2}) --- and analyzes the tradeoff between implementation simplicity and scalability.

\subsection{Single-Writer Model}
\label{section-single-writer}

A single privileged entity---either a user-space daemon or a kernel module, hereafter server---mediates all metadata access, centralizing both planes under one address space.
Because a single writer owns both planes, consistency is structural: no cross-process synchronization protocol is needed, and the implementation reduces to straightforward locking on internal data structures (R1, R2, R4).
Internally, the server may use fine-grained locking to allow concurrency among non-conflicting operations (e.g., concurrent receives on the same topic; \cref{section-concurrency}), while still guaranteeing atomicity where it matters---between membership changes and data-plane operations. Process exit detection and metadata cleanup complete in one place; on publisher crash, the server takes over Transient Local retention, enabling uninterrupted history delivery to late joiners. A kernel module additionally eliminates the intermediate scheduling layers present in daemon-based designs, structurally precluding priority misalignment from intermediate threads (\cref{section-discussion}). The tradeoff is a potential scalability bottleneck from the single update point (R5); \cref{section-evaluation} quantifies this cost.

\subsection{Owner-Driven Reclaim Model}
\label{section-owner-driven}

Each publisher manages the data plane for its own published
messages, as in iceoryx2's fully decentralized architecture.
Publishers maintain the global reference set in locally accessible
metadata (e.g., embedded in chunk headers), and subscribers update
it directly via shared memory, so reclamation decisions are made
locally without centralized coordination.
The control plane, however, cannot be localized: membership
information spans multiple processes and must be shared via
cross-process synchronization.
Cross-process mutexes on shared memory are the straightforward
option, but a process crashing while holding the mutex leaves it
permanently locked; POSIX robust mutexes
(\texttt{PTHREAD\_MUTEX\_ROBUST}) \cite{posix2017robust} handle
this case, but recovering data structure consistency after a crash
remains the application's responsibility, adding significant
implementation complexity.
Lock-free alternatives avoid the crashed-lock problem but
introduce their own verification challenges.
The scalability advantage is that different publishers operate on
disjoint data plane metadata, enabling parallel reclamation without
the single-writer bottleneck.

\subsection{Tradeoffs from Data plane/Control Plane Separation}
\label{section-tradeoffs}

In owner-driven reclaim, reclamation requires jointly reading the data plane and the separately managed control plane.
Without atomic access across both planes, race conditions arise.
For example, while a publisher verifies that all references to a message have been released, a subscriber may crash and a new subscriber may join concurrently, acquiring a reference via Transient Local history delivery.
Because the data plane read and membership update are not atomic, the publisher may observe neither the orphaned nor the new reference, and incorrectly reclaim it.
The same race class applies to publisher crashes: the recovery mechanism may free objects while a late joiner concurrently acquires references.
A publisher crash further causes a service continuity gap --- retention responsibility belongs to the crashed publisher, making Transient Local history permanently unavailable to late joiners, whereas single-writer transfers retention to the server.

Mitigating each race requires handshake protocols whose intermediate states interact with all other concurrent operations.
\cref{tab-concurrent-ops} enumerates all nine metadata-modifying operations.
Six of nine operations touch both planes; any concurrent pair among them is a potential race site.
The state space grows combinatorially, making exhaustive correctness verification increasingly difficult with each added protocol.
iceoryx2's development history illustrates this complexity: v0.3.0 (April 2024) fixed race conditions arising under frequent endpoint connection and disconnection \cite{iceoryx2_030_release}, and decentralized cleanup of resources left by crashed processes --- tracked as a separate effort \cite{iceoryx2_issue_96} --- required health monitoring infrastructure that arrived in v0.5.0 (December 2024) \cite{eltzschig2024iceoryx2v050}.

\begin{table}[tb]
\centering
\caption{Metadata-modifying operations in owner-driven reclaim:
lifecycle events for each endpoint type (publisher or subscriber), data-path operations,
and subscriber reference release.
Operations accessing both planes are potential race sites under
separate management.}
\label{tab-concurrent-ops}
\setlength{\tabcolsep}{4pt}
\begin{tabular}{@{}clcc@{}}
\toprule
\# & Operation & Data plane & Control plane \\
\midrule
1 & Subscriber join (incl.\ history delivery) & \checkmark & \checkmark \\
2 & Subscriber leave (graceful) & \checkmark & \checkmark \\
3 & Subscriber crash cleanup & \checkmark & \checkmark \\
4 & Publisher join & & \checkmark \\
5 & Publisher leave (graceful) & \checkmark & \checkmark \\
6 & Publisher crash cleanup & \checkmark & \checkmark \\
7 & Publish \footnotemark & \checkmark & \\
8 & Reclamation check & \checkmark & \checkmark \\
9 & Subscriber reference release & \checkmark & \\
\bottomrule
\end{tabular}
\vspace{-4mm}
\end{table}

\footnotetext{Publish accesses only the data plane assuming the
publisher caches the subscriber list locally; the cache is
refreshed during membership operations, not on every publish.
Stale caches are themselves a source of races in owner-driven
reclaim.}

In single-writer, both planes are managed by the same entity,
structurally precluding the cross-plane races without
dedicated protocol design.
\cref{section-evaluation} evaluates whether the resulting
scalability cost is acceptable for realistic workloads.

\section{Implementation in Agnocast}
\label{section-implementation}
This section implements \SMARTPOINTER{} (\cref{section-ipc-shared-ptr}) on Agnocast v2.3.0 \cite{agnocast230} \textsuperscript{\ref{fn:agnocast230}}, a true zero-copy IPC middleware for ROS~2, demonstrating that the proposed abstraction is realizable on a concrete pub/sub system.
Agnocast comprises a user-space library (\emph{agnocastlib}), a Linux kernel module (\emph{agnocast\_kmod}), and a heap-redirection hook injected via \texttt{LD\_PRELOAD} that redirects heap allocations between \lstinline|loan| and \lstinline|publish| to a per-process shared memory region (see Fig. 4 in \cite{agnocast2025} for the software stack).
Here, \lstinline|loan()| is a \lstinline|malloc()| call, and the hook redirects it---along with all heap allocations up to \lstinline|publish()|---to shared memory; \lstinline|publish()| then performs the metadata update and subscriber notification (\cref{section-ioctl}).
Topic discovery is handled internally by the kernel module;
interoperability with standard ROS~2 (DDS-based) communication
is provided by a separate bridge, described in~\cite{agnocast2025}.
The kernel module serves as the single-writer
(\cref{section-single-writer}): all data plane and control plane
metadata is managed within the module, accessed exclusively
through \texttt{ioctl} calls.

\vspace{-1mm}
\subsection{Kernel Module Data Structures}
\label{section-kmod-ds}
\vspace{-1mm}

The module maintains a per-topic data structure protected by a per-topic reader-writer lock; a global reader-writer lock guards the top-level hash table mapping topic names to per-topic structures (\cref{fig-kmod-data-structure}).
Following the single-writer model (\cref{section-single-writer}), both planes (\cref{section-system-model}) are merged: the entry tree realizes the data plane, while the endpoint tables and hash table realize the control plane.
Each per-topic structure contains the following:

\textbf{Entry tree.}
A red-black tree of entry records (one per published message) keyed by a monotonically increasing entry ID.
Each entry record stores the message's virtual address and a subscriber bitmap (\lstinline|DECLARE_BITMAP|); these two fields constitute the per-message metadata referenced in \cref{section-problem-statement}--\cref{section-metadata-architecture}.
This bitmap is the concrete realization of the global reference count (\cref{section-definition}): bit~$i$ is set if and only if the subscriber with topic-local ID~$i$ currently holds at least one \SMARTPOINTER{} reference.
A zero bitmap, combined with QoS Depth satisfaction, is the necessary and sufficient condition for reclamation (R1).

\textbf{Endpoint tables.}
Per-topic hash tables store publisher and subscriber records,
each containing a topic-local ID (determining the bit position
in every bitmap of that topic), owning PID, and QoS parameters.
Subscriber records additionally store a watermark for Transient Local history delivery.

\vspace{-5mm}
\subsection{Core Operations}
\label{section-ioctl}

\begin{figure}[tb]
  \centering
  \includegraphics[width=\linewidth]{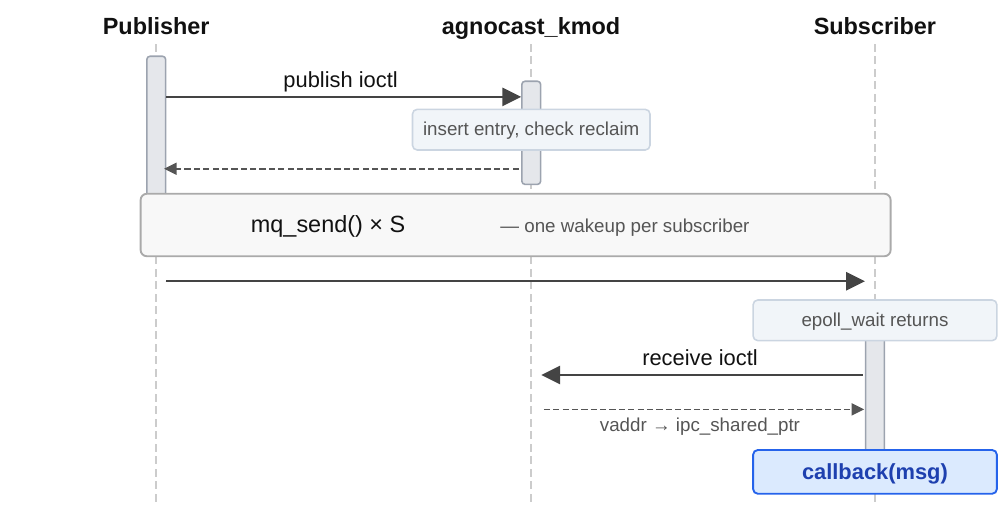}
  \vspace{-5mm}
 \caption{E2E message flow in Agnocast. The subscriber also issues a release \lstinline|ioctl| on each $1 \to 0$ local refcount transition.}
  \label{fig-e2e-message-flow}
  \vspace{-5mm}
\end{figure}

\begin{figure}[tb]
  \centering
  \includegraphics[width=\linewidth]{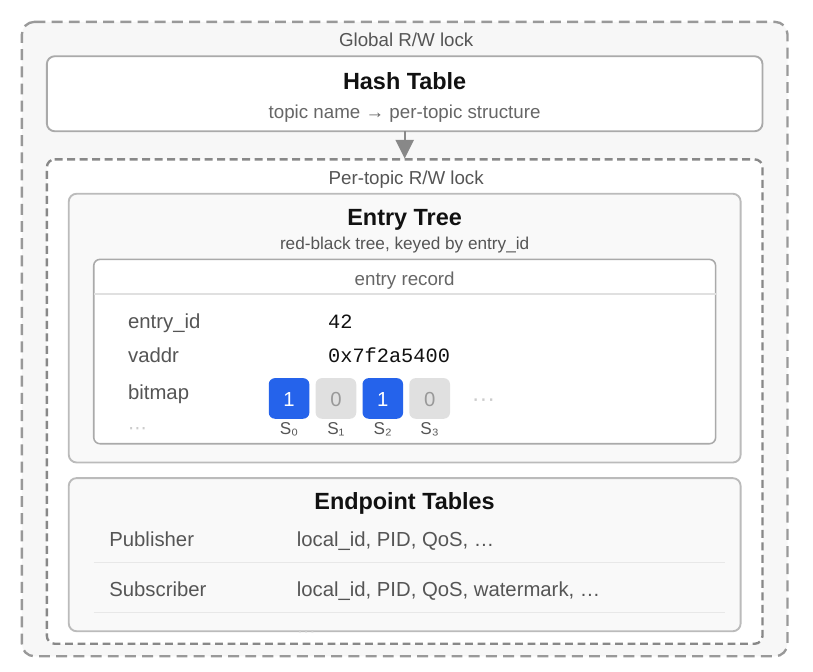}
  \vspace{-5mm}
  \caption{Kernel module data structure layout in Agnocast. Bit~$i$ of each
  entry's bitmap is set while subscriber~$i$ holds at least one
  reference; a zero bitmap with QoS depth satisfied triggers
  reclamation. The subscriber table stores a per-subscriber
  watermark for Transient Local history delivery.}
  \label{fig-kmod-data-structure}
  \vspace{-5mm}
\end{figure}

This subsection traces how each \SMARTPOINTER{} operation
(\cref{section-definition}) maps to a kernel \texttt{ioctl} (\cref{fig-e2e-message-flow}).

\textbf{Publish.}
The \texttt{ioctl} inserts a new entry record with a
zero-initialized bitmap, then scans the publisher's oldest
entries and evicts those whose bitmap is all-zero and that exceed
QoS Depth, returning their virtual addresses to user space for
deallocation.
This realizes the reclaimable timing of
\cref{section-definition}, including Transient Local retention
(R1).
The \texttt{ioctl} also returns the list of current subscriber
IDs; user space then sends a zero-length wakeup to each via a
per-subscriber POSIX message queue (capacity~1, registered with
\texttt{epoll}).
If the queue is full, the send returns immediately; the
subscriber is already scheduled to wake and its next receive
drains all pending entries.
This $O(S)$ post-publish notification is the dominant cost
measured in \cref{section-evaluation}.

\textbf{Receive.}
After wakeup, the subscriber issues this \texttt{ioctl}.
The module computes entries since the subscriber's watermark and, for each, atomically sets the subscriber's bit in the global reference count bitmap and returns the entry ID and virtual address.
On return, \emph{agnocastlib} constructs an \SMARTPOINTER{} for each entry with local reference count~1 (R3).

\textbf{Reference release.}
When an \SMARTPOINTER{}'s local count reaches zero,
\emph{agnocastlib} issues this \texttt{ioctl}, which atomically
clears the subscriber's bit in the bitmap.
No deallocation occurs; the entry becomes eligible for eviction
the next time the publisher publishes.
Only the local-count 0$\leftrightarrow$1 transition triggers a
kernel call; in-process copies modify only the process-local
count (R3).

\textbf{Endpoint registration.}
Registration inserts a new endpoint record with a fresh topic-local ID.
For Transient Local subscribers, the watermark is initialized to deliver the latest Depth retained entries on the first receive (R1).
Publisher-side QoS Depth bounds the number of entries retained in the entry tree for that publisher; subscriber-side QoS Depth determines the watermark's initial offset (how far back a late joiner begins receiving), but does not affect reclamation eligibility.

\subsection{Concurrency Control and Scalability}
\label{section-concurrency}

The module enforces a strict two-level lock hierarchy: a global
reader-writer lock over a per-topic reader-writer lock.
\cref{tab-lock-modes} shows the lock mode per operation.

\textbf{Per-topic parallelism (R5, scalability).}
Publish, receive, and release hold only a read lock on the
global lock; with $T$ independent topics, up to $T$ publish
critical sections execute concurrently, confining the
single-writer bottleneck to intra-topic serialization.

The receive and release paths use read locks on the per-topic lock as well: the receive path's writes are either atomic (\lstinline|test_and_set_bit| on the subscriber bitmap) or per-subscriber (watermark, mmap flag), with at most one receive per subscriber in flight at a time (reentrant callback groups require separate handling).
This allows all $S$ subscribers on the same topic to receive
concurrently, eliminating intra-topic serialization on the
receive side.
Only publish takes the per-topic write lock, as it modifies the
shared entry tree (insertion and eviction).
\cref{section-evaluation} quantifies this scalability.

\begin{table}[h]
\centering
\caption{Lock acquisition modes per operation.}
\label{tab-lock-modes}
\setlength{\tabcolsep}{4pt}
\begin{tabular}{@{}lcc@{}}
\toprule
Operation & Global lock & Per-topic lock \\
\midrule
Publish         & READ  & WRITE \\
Receive         & READ  & READ  \\
Release         & READ  & READ  \\
Membership change & WRITE & ---   \\
\bottomrule
\end{tabular}
\end{table}

\textbf{Structural atomicity (R4, dynamic membership).}
Membership changes hold the global write lock, serializing
against all concurrent operations.
No handshake protocol is required: the single lock structurally
guarantees the consistency that owner-driven reclaim must achieve
through cross-plane synchronization
(\cref{section-tradeoffs}).

\textbf{Process exit handling.}
The module detects process exit via a kernel tracepoint and
performs cleanup under the same global write lock: orphaned
subscriber bits are cleared from every bitmap, and reclaimable
entries are evicted (R2).
Because the server survives publisher crashes, Transient Local
history remains available to late joiners (R4) --- a property
unachievable in owner-driven reclaim where retention
responsibility belongs to the crashed publisher
(\cref{section-tradeoffs}).

\section{Evaluation} \label{section-evaluation}

\begin{figure*}[t]
  \centering
  \begin{minipage}[b]{0.32\linewidth}
    \centering
    \includegraphics[width=\linewidth]{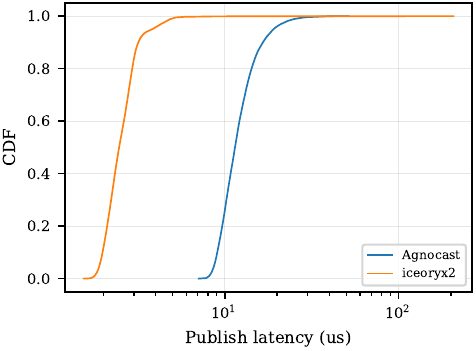}
  \end{minipage}
  \hfill
  \begin{minipage}[b]{0.32\linewidth}
    \centering
    \includegraphics[width=\linewidth]{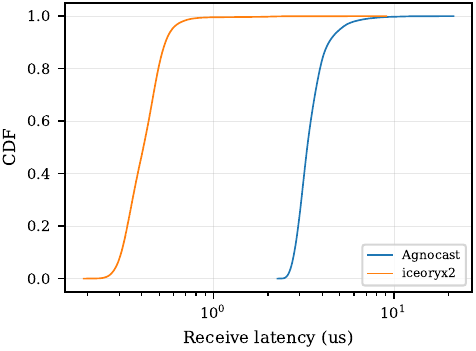}
  \end{minipage}
  \hfill
  \begin{minipage}[b]{0.32\linewidth}
    \centering
    \includegraphics[width=\linewidth]{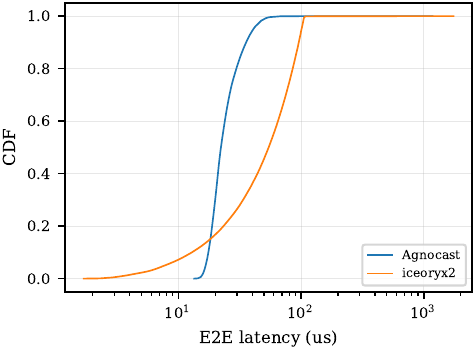}
  \end{minipage}
  \caption{Publish, receive, and E2E latency for a representative configuration ($T = 10, S = 4, R = 100$ Hz).}
  \label{fig-latency-comparison}
  \vspace{-4mm}
\end{figure*}

This section evaluates single-writer scalability for realistic robotics workloads.\footnote{Benchmark code, scripts, and reproduction pipeline (Agnocast branch \texttt{2.3.0-isorc26}): \url{https://github.com/sykwer/isorc26_benchmark}.}
As shown in \cref{section-birrell-relationship}, \SMARTPOINTER{} absorbs the dirty/clean RPCs of Birrell's reference listing~\cite{birrell1993distributed} into existing publish and membership operations, yielding a clear per-message communication advantage analytically; we therefore focus the empirical evaluation on the scalability comparison with iceoryx2's owner-driven reclaim.

\subsection{Experimental Design}
\label{section-experimental-design}

\textbf{Goal and workload baseline.}
We evaluate scalability of single-writer against iceoryx2's owner-driven reclaim across topic count $T$, subscriber fan-out $S$, and their combination (publish rate $R$ fixed; \cref{tab-sweeps}).
Autoware v1.7.1 (\cref{section-background}) provides the reference scale: 227 topics, median $S{=}1$ (mean 3.4), with a few high fan-out topics up to $S{=}35$.
Results defensible at this scale justify single-writer for realistic ROS~2 workloads.
In Autoware's typical pipeline, callback processing time ranges from several ms to tens of ms --- one to two orders of magnitude larger than IPC latency --- so metadata management overhead on the order of $\mu$s does not substantively impact end-to-end system performance.

\textbf{Systems and environment.}
We compare Agnocast (v2.3.0)\footnote{\label{fn:agnocast230}The evaluated build includes one post-release change: the per-topic lock for receive ioctl is relaxed from write to read, enabling concurrent receives on the same topic (\cref{tab-lock-modes,section-concurrency}). This change will be included in a future Agnocast release.} \cite{agnocast230} against iceoryx2 (v0.8.1) \cite{iceoryx2} on an Intel Xeon E-2278GE (3.30~GHz, 8~cores/16~threads), 32~GB RAM, Ubuntu~22.04, Linux~6.8, ROS~2~Humble, with all processes under FIFO real-time scheduling at priority~80.
To prevent Linux's default RT bandwidth throttle (a 50~ms enforced gap per 1~s window) from contaminating tail latency, we set \texttt{kernel.sched\_rt\_runtime\_us=999500}.
The two systems differ in notification mechanism: Agnocast sends a POSIX MQ message per subscriber after the publish \texttt{ioctl}, waking \texttt{epoll}-based event loops; iceoryx2 subscribers poll with \lstinline|receive()| at 100~$\mu$s intervals.
At $R{=}100$~Hz, this polling interval is small relative to the message period and not dominant in the results.

\textbf{Benchmark design.}
A one-publisher-per-topic architecture assigns independent pub/sub processes per topic ($T$ publishers $+$ $T{\times}S$ subscribers).
We use a fixed-size ${\approx}1$~KB message; since both systems perform true zero-copy IPC, message size does not affect latency.
The fixed size is imposed by iceoryx2's pool-allocation model, which requires self-contained payloads; Agnocast natively supports arbitrary types, making this benchmark conservative for Agnocast.
Each configuration runs for 5 iterations with a 2~s warmup, during which all membership changes complete, followed by 10~s of measurement.
Reported p50 and p99.9 are exact percentiles over the full sample set for each configuration (pooled over all iterations and per-process streams), not averages of per-iteration or per-process percentiles.
At large process counts, OS scheduling latency dominates metadata overhead and masks middleware-specific scalability.
We therefore bound the parameter space to keep scheduling effects secondary: on this machine, event-processing capacity is empirically ${\approx}140{,}000$~events/s, and all configurations are capped at 60\% utilization (84,000~events/s).
The sweep dimensions and measured metrics are summarized in \cref{tab-sweeps} and \cref{tab-metrics}, respectively.

\begin{table}[t]
\centering
\caption{Sweep parameters ($R = 100$~Hz throughout).}
\label{tab-sweeps}
\setlength{\tabcolsep}{5pt}
\begin{tabular}{@{}lllc@{}}
\toprule
Sweep & Variable & Range & Fixed \\
\midrule
A & $T$ (topics)       & 1--200                       & $S{=}2$ \\
B & $S$ (subscribers)  & 1--32                        & $T{=}10$ \\
C & $T \times S$       & $T$: 10--100,\ $S$: 2--16   & --- \\
\bottomrule
\end{tabular}
\end{table}

\begin{table}[t]
\centering
\caption{Measured metrics. $t_{\text{publish}}$ denotes the start of \lstinline|publish()|; $t_{\text{receive}}$ denotes the start of the subscriber callback. E2E latency excludes callback execution time.}
\label{tab-metrics}
\setlength{\tabcolsep}{4pt}
\begin{tabular}{@{}lp{6cm}@{}}
\toprule
Metric & Definition \\
\midrule
Publish latency &
  Wall-clock time from \lstinline|loan()| through
  \lstinline|publish()| completion, including shared memory
  allocation, payload construction, publish ioctl, and subscriber
  notification. \\[4pt]
Receive latency &
  Wall-clock time of the core receive-side ioctl
  (receive ioctl for Agnocast; \lstinline|receive()| for iceoryx2). \\[4pt]
E2E latency &
  $t_{\text{receive}} - t_{\text{publish}}$: full one-way
  delivery latency. \\
\bottomrule
\end{tabular}
\vspace{-5mm}
\end{table}

\subsection{Results}
\label{section-results}

\begin{figure*}[t]
  \centering
  \begin{minipage}[b]{0.32\linewidth}
    \centering
    \includegraphics[width=\linewidth]{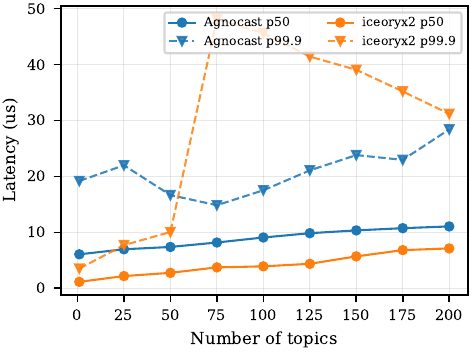}
  \end{minipage} \hfill
  \begin{minipage}[b]{0.32\linewidth}
    \centering
    \includegraphics[width=\linewidth]{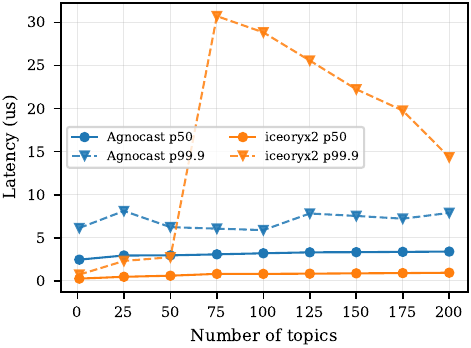}
  \end{minipage} \hfill
  \begin{minipage}[b]{0.32\linewidth}
    \centering
    \includegraphics[width=\linewidth]{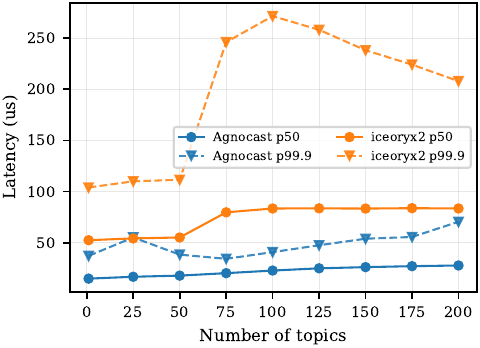}
  \end{minipage}

  \vspace{2mm}

  \begin{minipage}[b]{0.32\linewidth}
    \centering
    \includegraphics[width=\linewidth]{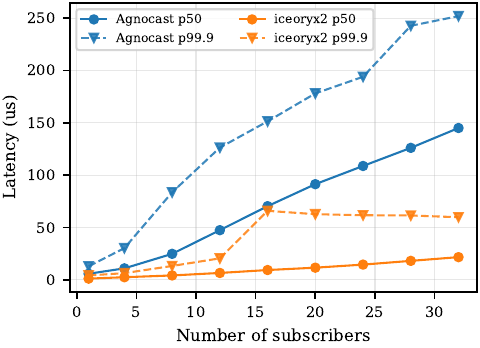}
  \end{minipage} \hfill
  \begin{minipage}[b]{0.32\linewidth}
    \centering
    \includegraphics[width=\linewidth]{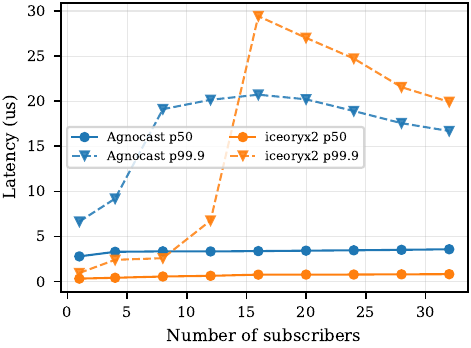}
  \end{minipage} \hfill
  \begin{minipage}[b]{0.32\linewidth}
    \centering
    \includegraphics[width=\linewidth]{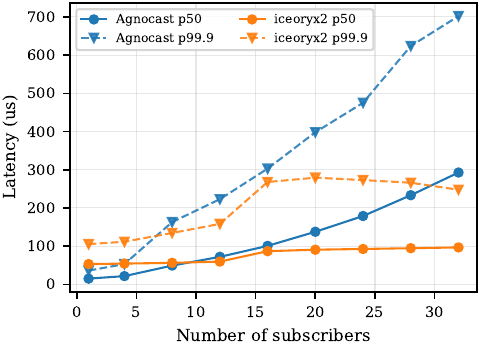}
  \end{minipage}

  \vspace{1mm}
  \caption{Latency scaling for publish (left), receive (center), and E2E (right) metrics as a function of the number of topics (top row, $S=2$) and the number of subscribers (bottom row, $T=10$).}
  \label{fig-latency-scaling-matrix}
  \vspace{-3mm}
\end{figure*}

\textbf{Per-operation latency distribution (\cref{fig-latency-comparison,tab-representative}).}
\cref{tab-representative} summarizes per-operation latencies at a
representative configuration ($T{=}10$, $S{=}4$, $R{=}100$~Hz).
iceoryx2's per-operation publish and receive latencies are
4--8$\times$ lower at p50, since both paths reduce to shared memory
operations whereas Agnocast incurs an ioctl and per-subscriber
POSIX MQ notification.
At the E2E level, however, Agnocast becomes \emph{lower} than
iceoryx2 by 2.5$\times$ at p50 (21.6 vs.\ 54.3~$\mu$s) and 2.1$\times$
at p99.9 (53.0 vs.\ 111.6~$\mu$s): iceoryx2's 100~$\mu$s polling
interval (\cref{section-experimental-design}) imposes a
delivery-latency floor that overwhelms its per-operation advantage
at this rate.

\begin{table}[t]
\centering
\caption{Per-operation latency at $T{=}10$, $S{=}4$, $R{=}100$~Hz.}
\label{tab-representative}
\setlength{\tabcolsep}{5pt}
\begin{tabular}{@{}lrrrr@{}}
\toprule
& \multicolumn{2}{c}{p50 ($\mu$s)} & \multicolumn{2}{c}{p99.9 ($\mu$s)} \\
\cmidrule(lr){2-3}\cmidrule(lr){4-5}
Metric & Agnocast & iceoryx2 & Agnocast & iceoryx2 \\
\midrule
Publish latency&  11.0 &  2.5 & 30.3 &   6.4 \\
Receive latency&   3.3 &  0.4 &  9.2 &   2.4 \\
E2E     latency&  21.6 & 54.3 & 53.0 & 111.6 \\
\bottomrule
\end{tabular}
\vspace{-3mm}
\end{table}

\textbf{Topic count scaling --- Sweep~A (\cref{fig-latency-scaling-matrix}, top row).}
Both systems' publish and receive latencies grow modestly with $T$.
Agnocast's publish p50 rises from 6.1~$\mu$s ($T{=}1$) to
11.1~$\mu$s ($T{=}200$) (1.8$\times$); receive p50 stays in
2.5--3.4~$\mu$s, confirming the receive ioctl is independent of $T$.
Agnocast's E2E p50 ranges 15--28~$\mu$s and p99.9 stays below
71~$\mu$s across all $T$.
iceoryx2's E2E p50 starts at 53~$\mu$s --- a floor set by its 100~$\mu$s
polling interval (\cref{section-experimental-design}) --- and grows
to 84~$\mu$s; p99.9 grows from 104~$\mu$s ($T{=}1$) to a peak of
272~$\mu$s at $T{=}100$, then settles to 208~$\mu$s at $T{=}200$.
At $T{=}200$, $S{=}2$, Agnocast's E2E p99.9 (71~$\mu$s) is 2.9$\times$
lower than iceoryx2's (208~$\mu$s) despite iceoryx2's faster
per-operation paths --- the polling-derived floor dominates
at this rate.

\textbf{Subscriber count scaling --- Sweep~B (\cref{fig-latency-scaling-matrix}, bottom row).}
Agnocast's publish p50 grows linearly from 5.8~$\mu$s ($S{=}1$) to
145.0~$\mu$s ($S{=}32$) --- a direct measurement of the $O(S)$ cost
of sending one POSIX MQ notification per subscriber per publish
(\cref{section-ioctl}).
iceoryx2's publish also scales with $S$, from 1.2 to 21.7~$\mu$s,
since each subscriber owns a separate shared-memory channel that
the publisher must enqueue into; however, each enqueue is a
\textit{userspace} memory operation, an order of magnitude cheaper
than Agnocast's per-subscriber syscall.
Agnocast's receive p99.9 stays bounded under 21~$\mu$s across the
entire sweep (peak 20.7~$\mu$s at $S{=}16$): the receive path
acquires only a per-topic read lock (\cref{tab-lock-modes}),
so all $S$ subscribers execute their receive ioctls concurrently
without serialization.
For E2E latency, Agnocast's p50 grows steeply from 15.0 to
292.7~$\mu$s (19.5$\times$), while iceoryx2's grows from 53.1
to 96.7~$\mu$s (1.8$\times$).
The two systems cross over near $S{=}8$ at p99.9 (Agnocast
163.1~$\mu$s vs.\ iceoryx2 134.5~$\mu$s): at $S{=}32$, Agnocast's
E2E p99.9 (702.1~$\mu$s) exceeds iceoryx2's (247.7~$\mu$s) by
2.8$\times$, a regime where the per-subscriber syscall cost of
notification overtakes iceoryx2's polling overhead.
Publish latency alone (145~$\mu$s) accounts for half of Agnocast's
E2E p50 at $S{=}32$, confirming that the bottleneck is the $O(S)$
notification mechanism, not the single-writer metadata architecture
(receive p99.9 is only 16.7~$\mu$s).

\begin{figure}[tb]
  \centering
  \includegraphics[width=\linewidth]{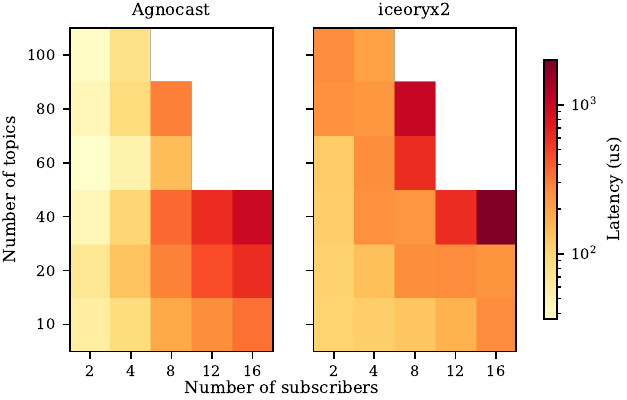}
  \vspace{-5mm}
  \caption{E2E p99.9 latency heatmaps over (T, S) parameter space.}
  \label{fig-component-oriented-realtime}
  \vspace{-5mm}
\end{figure}

\textbf{Combined scaling --- Sweep~C (\cref{fig-component-oriented-realtime}).}
Agnocast's hot zone runs along the high-subscriber edge:
($T{=}40$, $S{=}16$) reaches 1{,}000~$\mu$s p99.9, driven by the
$O(S)$ notification cost; the high-topic low-fan-out region
remains cool ($T{=}100$, $S{=}2$: 39~$\mu$s).
iceoryx2's hot zone is pushed further along the same edge:
($T{=}40$, $S{=}16$) reaches 1{,}992~$\mu$s p99.9, driven by
polling contention at high subscriber-process counts;
($T{=}100$, $S{=}2$): 269~$\mu$s.
Agnocast retains the p99.9 advantage across most of the grid;
iceoryx2 is faster only in a narrow band ($S{=}8$--$12$
at $T{\leq}40$), where Agnocast's $O(S)$ syscall cost has
grown but iceoryx2's polling cost has not yet collapsed.

\subsection{Discussion}
\label{section-discussion}

\textbf{Sufficiency for the Autoware workload.}
At Autoware-scale topic counts and typical fan-out 
(median S=1, mean S=3.4), Agnocast's E2E p99.9 stays 
on the order of tens of microseconds --- one to two 
orders of magnitude below Autoware's callback processing 
time --- and is consistently lower than iceoryx2's. 
High fan-out topics are the greater challenge: at 
$S \approx 35$ (the most subscribed Autoware topic), 
linear extrapolation from Sweep B places Agnocast's 
E2E p99.9 on the order of hundreds of microseconds, 
exceeding iceoryx2 in this regime. Three factors 
mitigate the impact: first, high fan-out topics are rare (only 5 of 227 Autoware topics have $S{\geq}16$); second, several of these are low-rate (HD map and static coordinate transforms are published only on change); third, the $O(S)$ cost originates from per-subscriber MQ notification, not the single-writer metadata architecture, and is targeted for a post-v2.3.0 release.

\noindent\textbf{Notification mechanism vs. metadata architecture.}
Both systems incur $O(S)$ publish cost, but the constant differs by an order of magnitude: Agnocast issues an MQ syscall per subscriber (${\approx}4.5$~$\mu$s/sub), whereas iceoryx2 performs a userspace shared-memory enqueue (${\approx}0.7$~$\mu$s/sub).
This per-subscriber syscall cost --- not the single-writer metadata architecture --- drives Agnocast's fan-out scaling regression.
Conversely, iceoryx2 incurs a 100~$\mu$s polling-interval floor on delivery latency, dominant at low-to-moderate fan-out where Agnocast's event-driven wakeup wins.
The receive-side read lock (\cref{tab-lock-modes}) ensures that the single-writer metadata path itself introduces no serialization among subscribers: at $S{=}32$, receive p99.9 is only 16.7~$\mu$s, comparable to iceoryx2's 19.9~$\mu$s.
The single-writer metadata architecture is compatible with either notification mechanism; the choice between event-driven and polling is an independent design axis whose tradeoff depends on workload.

\textbf{Syscall overhead.}
The per-operation latency gap (\cref{tab-representative}; p99.9 publish: $4.7{\times}$, receive: $3.8{\times}$) reflects the user-kernel transition cost of implementing single-writer as a kernel module.
At the E2E level, however, Agnocast is \emph{lower} than iceoryx2 by 2.5$\times$ at p50 and 2.1$\times$ at p99.9, confirming that \texttt{ioctl} overhead is dominated by other E2E terms (notification path and polling-interval floor).
Kernel module placement also enables kernel-level process exit detection and atomic metadata cleanup (\cref{section-concurrency}).
It further eliminates the intermediate kernel threads that propagate priority misalignment in daemon-based IPC \cite{kim2025cros} (\cref{section-related-work}).

\section{Related Work} \label{section-related-work}

\textbf{Distributed reference counting.}
\SMARTPOINTER{} is structurally closest to Birrell et al.'s reference listing~\cite{birrell1993distributed} among distributed reference counting methods, adapted and specialized for the pub/sub domain.
Weighted reference counting~\cite{bevan1987distributed, watson1987efficient} established the field, and many variants followed~\cite{piquer1991indirect, goldberg1989generations}; see~\cite{plainfosse1995survey, abdullahi1998garbage} for surveys.
Other safe memory reclamation schemes such as hazard pointers~\cite{michael2004hazard}, epoch-based reclamation~\cite{fraser2004practical}, and RCU~\cite{mckenney1998read} enable lock-free reader access to concurrent data structures primarily within a single address space, and do not consider publisher/subscriber membership or QoS-driven retention.
In contrast, \SMARTPOINTER{} specializes Birrell's structure for the pub/sub domain by exploiting that reference destinations are derivable a priori from membership, that references flow unidirectionally from publishers to subscribers, and that messages form a cycle-free object graph; the structural relationship is detailed in \cref{section-birrell-relationship}.

\textbf{True zero-copy IPC for ROS~2.}
TZC \cite{wang2019tzc} achieves zero-copy for dynamically sized arrays via partial serialization but requires message-type-specific implementations and pre-determined sizes; LOT \cite{iordache2021lot} uses \texttt{boost::interprocess} for ROS~1 but requires library-specific message formats.
Both impose substantial application-code constraints, impractical for large-scale ROS~2 projects.
iceoryx1 \cite{iceoryx1} centralized the control plane via a RouDi daemon; iceoryx2 \cite{iceoryx2} eliminated it.
Both restrict messages to self-contained, statically sized types; the race conditions arising from iceoryx2's decentralization are analyzed in \cref{section-tradeoffs}.
Fast DDS and Cyclone DDS provide shared memory transports but incur serialization costs and do not support true zero-copy for unsized types; Cyclone DDS integrates iceoryx1 as its backend \cite{cyclonedds_iceoryx}, inheriting the fixed-size restriction.
Agnocast \cite{agnocast2025} sidesteps the allocator constraint by mapping the publisher's heap to shared memory; this paper addresses its cross-process lifetime management.
CROS-RT \cite{kim2025cros} addresses priority inversion in DDS-based ROS~2 IPC by propagating priorities across application, middleware, and kernel layers, the root cause being intermediate kernel threads interposed by the DDS transport.
Agnocast's kernel module structurally eliminates this: \texttt{ioctl}-based interaction bypasses such threads, leaving no propagation path for priority misalignment.
A user-space daemon would reintroduce them on every metadata update, distinguishing the kernel module choice from daemon-based alternatives beyond the fault-tolerance argument of \cref{section-concurrency}.
Luo et al.\ \cite{luo2025flexible} propose a multi-stage zero-copy approach enabling end-to-end zero-copy across processing chains.
They note that, at the time of \cite{luo2025flexible}, Agnocast's custom executor precluded native ROS~2 components such as \texttt{message\_filter}, forcing non-zero-copy fallback; Agnocast v2.3.0 provides \texttt{message\_filter}-equivalent functionality within the executor, eliminating this disadvantage.

\section{Conclusion} \label{section-conclusion}

We proposed \SMARTPOINTER{}, a pub/sub-aware two-level reference
counting abstraction bounding global metadata updates to
per-subscriber $0\leftrightarrow{}1$ transitions.
The key design tradeoff --- implementation simplicity versus
scalability --- favors single-writer: centralizing all metadata
access through a single writer provides structural atomicity across
the data plane and control plane, eliminating the cross-plane races
that demand complex handshake protocols in owner-driven reclaim.
Comparative evaluation against iceoryx2 at Autoware-class workload
scale demonstrates that the resulting scalability cost is acceptable,
confirming implementation simplicity as the rational choice for
realistic robotics workloads.

\section{Acknowledgment}
The authors thank all contributors to the development of Agnocast, and in particular Ryuta Kambe and Yutaro Kobayashi for their significant contributions.
This research is based on results obtained from a project, Green Innovation Fund Projects / Development of In-vehicle Computing and Simulation Technology for Energy Saving in Electric Vehicles (JPNP21027), subsidized by the New Energy and Industrial Technology Development Organization (NEDO).

\bibliographystyle{IEEEtran}
\bibliography{references}

\end{document}